\documentclass[11pt,romanappendices,conference,twocolumn]{IEEEtran}
\usepackage{cite}
\usepackage{algorithm2e}
\usepackage{amssymb}
\usepackage{enumerate}
\usepackage{graphicx}
\usepackage{amssymb}
\usepackage{amsbsy}
\usepackage{float}
\usepackage{balance}
\usepackage{url}
\usepackage{array}
\usepackage{varioref}
\usepackage[normalem]{ulem}
\usepackage{color}
\usepackage{empheq}
\usepackage{flushend}
\begin{document}

\bstctlcite{asms:BSTcontrol}
\title{Cellular-Broadcast Service Convergence \\through Caching for CoMP Cloud RANs}
\author{\IEEEauthorblockN{Symeon Chatzinotas, Dimitrios Christopoulos, Bj\"{o}rn Ottersten}
\IEEEauthorblockA{SnT - securityandtrust.lu,  University of Luxembourg,
\\Email: \textbraceleft Symeon.Chatzinotas, Dimitrios.Christopoulos, Bjorn.Ottersten\textbraceright@uni.lu}
}
\maketitle
\begin{abstract}
 Cellular and Broadcast services have been traditionally treated independently due to the different market requirements, thus resulting in different business models and orthogonal frequency allocations. However, with the advent of cheap memory and smart caching, this traditional paradigm can converge into a single system which can provide both services in an efficient manner. This paper focuses on multimedia delivery through an integrated network, including both a cellular (also known as unicast or broadband) and a broadcast last mile operating over shared
spectrum. The subscribers of the network are equipped with a cache which can effectively create zero perceived latency for multimedia delivery, assuming that the content has been proactively and intelligently cached. The main objective of this work is to establish analytically the optimal content popularity threshold, based on a intuitive cost function. In other words, the aim is to derive which content should be broadcasted and  which content should be unicasted.  To facilitate this, Cooperative Multi-Point (CoMP) joint processing algorithms are employed for the uni and broad-cast PHY transmissions. To practically implement this, the integrated network controller is assumed to have access to traffic statistics in terms of content popularity. Simulation results are provided to assess the gain in terms of total spectral efficiency. A conventional system, where the two networks operate independently, is used as benchmark.
 \end{abstract}
\section{Introduction}
\subsection{Where it all began}
In the history of wireless communications, there has been a number of services which have met commercial success and have driven not only the deployment of wireless infrastructure but also future evolutions in this domain. One of the first such services was audio and video broadcasting. The combination of low carrier frequencies with high penetration and the fact that a large number of subscribers can be simultaneously served led to ubiquitous adoption of broadcasting services. As a result, a wide range of prime spectrum is allocated to broadcasting. A proportionally sized  infrastructure, in form of radio towers and user equipment,  is currently deployed worldwide.

In parallel, the past three decades a new wireless service has been constantly expanding as a response to the demand for interactive services. Although this service has evolved through a series of generations, the term cellular will be used herein as a reference name. The cellular service was originally established for bidirectional voice  communications. However,  during the last decade,  the rapid expansion of all-IP communications is becoming the primary service delivered by cellular networks, by aggregating voice and data/Internet communications under the same umbrella. Due to the wide range of services and the advent of smartphones, cellular has met an unprecedented adoption and the design of its fifth generation network (5G) is underway. An expanding base of allocated spectrum is a substantial enabler of this upcoming new generation of cellular systems.

\begin{figure}[h]
\centering
\includegraphics[width=0.9\columnwidth]{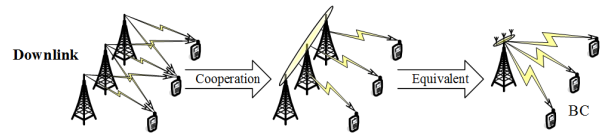}
\caption{Downlink base-station cooperation   of a CoMP network.}\label{fig: DL/UP CoMP}
\end{figure}

A key role for 5G systems is reserved by multi-antenna wireless communications. The cooperation of multiple antennas is admitted as the most prominent way of optimizing, in some required sense, the PHY layer of wireless systems. An example of this basic concept is given in Fig. \ref{fig: DL/UP CoMP}. An added benefit of this architecture lies in the inherent flexibility to operate the multi-antenna transmitters in either unicast or broadcast mode, which enables the considerations hereafter presented.

A review of technical and architectural solutions that have realistic possibility to achieve the targets of future wireless access set in 5G can be found  in \cite{Zander2013}. Therein, it was argued that although further improvements
in the PHY layer are expected, it is
unlikely that this alone could provide for the projected 1000-fold
increase in data traffic between 2010 and 2020. In light of this conjecture, the present work aims at establishing viable and cost efficient solutions for the next generation of wireless communication systems.
\subsection{Cloud Radio Access Networks}
\begin{figure}[h]
\centering
\includegraphics[width=0.95\columnwidth]{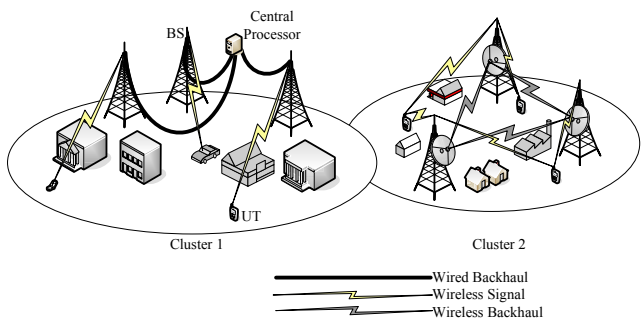}
\caption{System-wide CSI and ideal backhauling to radio heads enables centralized PHY processing.}
 \label{fig:  BScoop}
\end{figure}

Included in the most disruptive technology directions for 5G \cite{Boccardi2014}, radio access networks organized in a cloud architecture
is a recent design paradigm \cite{Meerja2015}. In the present section, we discuss a number of issues that can determine the architecture of Cloud Radio Access Networks (CRAN)  \cite{CRAN2011} and highlight the main barriers and opportunities.
An example of a CRAN network is given in Fig. \ref{fig:  BScoop}.
 \subsubsection{Network Neutrality and Business Models}  Much of the Internet success actually resulted from the fact that the network is agnostic in terms of the content that it carries flexibility. On the downside,  network operators have to support data hungry users without generating additional revenue. This situation has lead to extremely high network throughput requirements for 5G, mainly generated by multimedia traffic combined with sub-optimal spectrum allocation.  Potential solutions include additional spectrum allocation, infrastructure densification and higher spectral efficiency.
\subsubsection{Additional Spectrum} In practice spectrum is very congested,  leaving an insufficient to satisfy the projected throughput increase in 5G amount of available frequencies. Particularly, spectrum from 15GHz and above,  is envisaged to provide short range, low complexity hot spot coverage especially for indoor environments, while lower frequencies are reserved for wider area,  cellular coverage \cite{dahlman20145g}. The worthiness of multicell coordination in indoor environments has been studied in \cite{Kang2012Globecom}.
\subsubsection{Infrastructure Densification} Another approach would be to enable denser spatial frequency reuse through the deployment of micro/femto cells. Nevertheless, this would entail a large investment from the network operators which do not share the additional revenue generated by a higher-throughput network. Furthermore, more aggressive frequency reuse in a variable-cell size network creates a range of intra-system interference issues which have to be tacked before deployment.
\subsubsection{Higher Spectral Efficiency} Improving the spectral efficiency is the holy grail of 5G. Nevertheless, Cellular systems of the previous four generations are already approaching the channel capacity bounds. Coordinated MultiPoint (CoMP), which enables full frequency reuse across the whole network seems to require furher resources to cover for the projected demands.

In this context, the main question is:
\textbf{How could CRANs in the context of 5G provide an even higher throughput?}
\subsection{Evolving Traffic Patterns \& Multimedia Delivery}
Numerous works have focused on analyzing the traffic demand on networks delivering internet traffic\cite{venmani2012demystifying,paul2011understanding}. In order to better understand the problem we have to delve into the patterns of current Internet traffic. Originally, Internet  traffic was heavily based on web browsing, which is bursty in nature and did not contain large multimedia files. However, nowadays more and more applications and services require streaming of large multimedia files and mainly video. This shift in traffic patterns is mainly motivated by the consumer habits, which have evolved from consuming linear broadcasted content to on-demand multimedia.  Based on recent statistics, see for instance \cite{forecast2013cisco}, more than half of the Internet downlink is dedicated to video streaming services, such as YouTube and NetFlix. More importantly, it seems that certain multimedia files are extremely popular and are downloaded by a large percentage of subscribers. The most prominent example of such content are the so-called viral videos. Such  traffic demand patterns can be captured by a popularity distribution function. The popularity distribution of available multimedia content exhibits heavy tails, meaning that a very small number of multimedia are almost globally requested and a very large number of multimedia are almost individually consumed\cite{breslau1999web}.
However, current cellular standards have  been designed for unicast services, such as web browsing, and cannot exploit these evolving traffic patterns. This entails that even highly popular content would have to be unicasted to each individual subscriber by treating the cellular network as a ``dump pipe". Based on these facts, the evolved multimedia broadcast/multicast service (eMBMS) has been included in LTE-advanced \cite{lecompte2012evolved}.
\subsection{Cellular-Broadcast Convergence through Caching }
In this section, we describe a possible solution to the aforementioned problem. The key point is that globally popular multimedia can be much more efficiently distributed through broadcasting services, since it is meant to reach a high percentage of the subscriber base. Nevertheless, the subscriber expects to consume this content on demand and this implies that content should be stored locally through intelligent caching algorithms.  Multimedia that are seldomly requested can be distributed through the usual unicasting-cellular services. In the following paragraphs, we develop these ideas by providing more details on specific aspects:

\subsubsection{Cellular-Broadcast Convergence} As mentioned before, these two modes of multimedia distribution currently form two independent networks, which follow their own standards and use separate frequency ranges. It is argued herein that there is much to be gained in terms of spectral efficiency by considering converging cellular and broadcast services. Firstly, the available spectrum can be flexibly integrated and optimized to match the evolving traffic patterns. Secondly, broadcasters and multimedia providers would be able to benefit from the cellular uplink in order to determine and follow the current traffic trends. Finally, the multimedia delivery could be optimized by selecting the cellular or broadcast mode in order to optimize the system throughput.

\subsubsection{Proactive Intelligent Caching} This feature is a highly critical part of the solution, since it enables the subscriber to access the requested multimedia with zero-perceived delay. Caching looks more feasible nowadays, since the memory cost per GB keeps dropping.  In this direction, including a large cache at the subscriber equipment is becoming a financially viable solution. At the same time, many works (eg. \cite{Borst2010cache} and the references therein)   have investigated intelligent caching algorithms which aim at optimizing the caching and updating actions based on a number of objectives, such as global/local popularity \cite{brinton2013intelligent}, cache/file size and temporal variations.

In this paper, we provide an analytical model for evaluating the system throughput gain of the 5G proposed architecture in comparison to conventional cellular or broadcast architectures. For the sake of fairness, this comparison considers the same amount of available bandwidth and radio towers and focuses solely on improving the spectral efficiency by minimizing the time required to deliver a fixed data volume to the subscribers.

The remainder of this paper is structured as follows: Section \ref{sec: system model} presents the system model while Sec. \ref{sec: traffic Model} describes the considered multimedia traffic model and investigates the optimization of mode selection. In Sec. \ref{sec: numerical results},   the effect of different traffic parameters is numerically evaluated and Sec. \ref{sec: conclusion} concludes the paper.


\section{Architecture \& PHY transmission}\label{sec: system model}
\begin{figure}
\centering
\includegraphics[width=0.9\columnwidth]{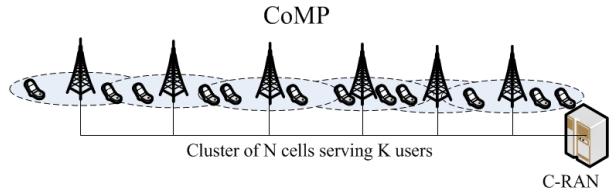}\\
\caption{Linear arrangement of a cellular system with \(N\) cooperating base stations and K users. }\label{fig:  system}
\end{figure}
Let us consider a cellular deployment of Base Stations (BSs), each being equipped with a single antenna. We focus on a cluster of \(N\) BSs which serve \(K\) single-antenna users in total with \(K>>N\). The cellular network can operate in two transmission modes: Unicast, where individual messages are transmitted to each user and Broadcast, where a single message is transmitted to all users. These two modes are used orthogonally, either in frequency or in time, so there is no intra-system interference between them. Furthermore, in order to optimize the spectral efficiency of the system, it is assumed that all BSs are connected to a central processor (e.g. C-RAN), which has access to both user requests and Channel State Information (CSI). This C-RAN is responsible for selecting the appropriate mode and calculating the signals to be transmitted by each BS.

To formally model our system, the received signal at the $i$-th user will read as $y_{i}= \mathbf h^{T}_{i}\mathbf x+n_{i},$
where \(\mathbf h^{T}_{i}\) is a \(1 \times N\) vector composed of the channel coefficients (i.e. channel gains and phases) between the \(i\)-th user and the  \(N\) BSs, \(\mathbf x\) is the \(N \times 1\)  vector of the transmitted symbols and  \(n_{i}\) is the complex circular symmetric (c.c.s.) independent identically distributed (i.i.d) zero mean  Additive White Gaussian Noise ($\mathrm{AWGN}$),  measured at the \(i\)-th user's receiver.
The general linear signal model in vector form reads as $ \mathbf y = \mathbf {H}\mathbf x + \mathbf n = \mathbf {H}\mathbf {W} s+ \mathbf{ n}$, where $ \mathbf {y \text{ and }  n } \in \mathcal{\mathbb{C}}^{K}$ and $\mathbf {x} \in \mathbb{C}^{N_{}}$. The channel matrix   $ \mathbf {H} \in \mathbb{C}^{K \times N}$ is composed of coefficients dependent on the assumed channel model. Let $\mathbf w_k \in \mathbb{C}^{N\times 1}$ denote the precoding weight vector applied to the Distributed Antenna System (DAS) to beamform towards all  users. The power radiated by each BS is  a  linear combination of all precoders \cite{Yu2007}, $ P_n = \left[\sum_{k=1}^N \mathbf w_k \mathbf w_k^\dag \right]_{nn}$, where $n$ is the antenna index.
\subsection{Operation Modes}
\subsubsection{Unicast Mode (UC)}
In this mode, the BS cluster transmits an individual stream to each user and thus it can support only \(N\) users simultaneously. In this context, the  weighted max-min fair SINR optimization problem  \cite{Christopoulos2014} with per-BS power constraint is defined as:
\begin{align}
&\mathcal{UC:}\    \max_{\  t, \ \{\mathbf w_k \}_{k=1}^{N}}  t& \notag\\
\mbox{s. t. }  &\frac{1}{\gamma_i}\frac{|\mathbf w_k^\dag \mathbf h_i|^2}{\sum_{{l\neq k }}|\mathbf w_l^\dag\mathbf h_i|^2+\sigma_i^2 }\geq t, &\label{const: F SINR}\\
&\forall i, l, k, \in\{1\dots N\},\notag\\
 \text{a.t. }\  & \left[\sum_{k=1}^N  \mathbf w_k\mathbf w_k^\dag  \right]_{nn}  \leq P_n, \forall n\in \{1\dots N\} \label{eq: max-min fair power const1}
 \end{align}
 where \(\gamma_i\) is the SINR of the \(i\)th user and $t$ is the optimization slack variable. The solution of this problem can provide the per-user spectral efficiency for the unicast mode $spf_\mathrm{uni}$. 

\subsubsection{Broadcast Mode (BC)}
In this mode, a single stream is transmitted to all \(K\) users and the caching algorithm of each user decides which content should be cached for future consumption based on the individual user requests.


If the practical but more elaborate per-BS power constraints are considered, the per-BS constrained, weighted, max-min fair, multicast  beamforming problem reads as
 \begin{align}
\mathcal{BC:}\    \max_{\  t, \ \mathbf w}  \ t& \notag\\
\mbox{subject to } & \frac{1}{\gamma_i}\frac{|\mathbf w^\dag \mathbf h_i|^2}{\sigma_i^2 }\geq t, \
i = 1\dots K,\label{const: F SINR}\\
 \text{and to }\ \ \ \ & \left[  \mathbf w\mathbf w^\dag  \right]_{nn}  \leq P_n,  \forall n\in \{1\dots N\},\label{eq: max-min fair power const2}
 \end{align}
 where $\mathbf w_k\in \mathbb{C}^{N}$ and $t \in \mathbb{R}^{+}$.
This problem is an instance of the weighted fair multigroup multicast beamforming problem of \cite{Christopoulos2014}, for one multicast group. More details for the solution of this problem are provided therein. The solution of this problem can provide the per-user spectral efficiency for the broadcast mode $spf_\mathrm{bc}$.
\section{Multimedia Traffic Model} \label{sec: traffic Model}
In order to study this problem, a traffic model based on the multimedia popularity is defined in this section. The
popularity is measured based on the number of requests and can be tractably described though a probability function. A widely-used abstraction for this function is the Zipf law, which is given by \cite{breslau1999web}
\begin{align}
f(i) =  \left(\frac{1}{i}\right)^a\label{eq: Zipf}
\end{align}
In more detail,
if we ordered the files from most to least popular at a given
point in time, then the relationship governing the frequency at
which the file of rank $i$ will appear is given by \eqref{eq: Zipf}. Consequently, the probability of a request occurring for file $i$ is
inversely proportional to its rank, with a shaping parameter
$\alpha$. A detailed analysis on how to choose the exact value of $\alpha$ is part of future extensions of this work \cite{hernandez2004variable}. For the sake of simplicity, it is assumed that all multimedia files have equal size and that there are no temporal variations of the popularity measure. In addition, we assume that the cache size is not a limiting factor in this study and relaxations of these constraints are part of future work.
\subsection{Threshold Optimization }
To easily present the novel concepts of the present work, the spectral efficiencies of the previous section will be employed as determinist parameters to assist the optimization process.
This can be achieved by defining a popularity threshold above which multimedia are broadcasted and below which multimedia are unicasted.
The decision on which mode to operate will be made based on the demand of the $i$-th file. In order to determine the optimal threshold above which all packets will be unicasted, let us proceed with the following assumptions. Each of the $K$ users requires only one file of size $s$ [bits] out of \(i_{max}\) files.  Let $i_{th}$ denote the optimal threshold and $spf_\mathrm{bc}$ and $spf_\mathrm{uni}$ denote the spectral efficiencies of the BC and the unicast configurations respectively, while $W$ denote the total available bandwidth.   Then the transmitted volume of data in BC mode will be given by $ V_\mathrm{bc}= (i_{th} -1)\cdot s $, since each file has to be broadcasted only once. The received/cached data volume through the BC mode across all \(K \) users will be $ s \cdot K  \cdot \sum_{i=1}^{i_{th}-1}f(i)$, because the function \(f\) determines the percentage of users that have requested the broadcasted files. On the other hand, in unicast-cellular mode, the transmitted volume of data will be  $V_\mathrm{uni}=s\cdot K  \cdot\sum_{i=i_{th}}^{i_{max}}f(i)$,  since each file is individually transmitted to each user that was requested it. As a result,  the received volume is by definition equal to the transmitted volume. Assuming a fixed operational bandwidth \(W\), the goal is to find the optimal threshold $i_{th}$ that minimizes the total transmission time\footnote{It should be noted that an equivalent formulation can be expressed in terms of fixing the required time and minimizing the required bandwidth.}, thus maximizing  the spectral efficiency of the total system:
\begin{align}
\mathcal{T:}\   & \min_{i}  \ T_\mathrm{tot} = T_\mathrm{uni} +T_\mathrm{bc}\label{eq: cost}
 \end{align}
  where $T_\mathrm{uni} = V_\mathrm{uni}/( W\cdot spf_\mathrm{uni})$ and $T_\mathrm{bc} = V_\mathrm{bc}/( W\cdot spf_\mathrm{bc})$. For $\alpha > 1$ in \eqref{eq: Zipf}, the optimization problem  \eqref{eq: cost} can be straightforwardly solved by derivation.  The optimal threshold between broadcasting and unicasting information in the cellular network is given by optimizing the following cost function
\begin{align}
T_{tot}(i) =  s/W\cdot\left(K\frac{\sum_{i}^{i_{\max}}f(i)}{C_{\max}spf_{\mathrm{uni}}}+\frac{i-1}{spf_{\mathrm{bc}}}\right)
\end{align}
where $ C_{max} = \sum_{i=1}^{i_{\max}}f(i) $ is a normalization parameter for the discretized pdf \(f\).  $T_{tot}$ is a linear function over $i$. Following tedious  algebraic derivations and by replacing the summations over discrete functions with integrals, i.e. $\sum_{i=1}^{i_{th}}f(i) = \int_{i=1}^{i_{max}}f(t)dt$, the optimal threshold between broadcasting and unicasting data is given by
\begin{align}
i_\mathrm{ th} = \left( \frac{{K}\cdot spf_\mathrm{bc}}{spf_\mathrm{uni}C_{max}+K\cdot spf_{bc}\cdot f(i_{\max})}\right)^\frac{1}{\alpha}\label{eq:  th}
 \end{align}

From \eqref{eq:  th} we can see that the optimal threshold between broadcasting and unicasting data depends on the spectral efficiencies of the two systems, the total number of users and the parameters of the traffic pdf. From this expression, some conclusions can be straightforwardly reached: 1) Increasing the number of system users or the BC spectral efficiency favors the BC mode, 2) Increasing the UC spectral efficiency or the shaping parameter favours the UC mode. The latter finding can be intuitively explained by the fact that higher shaping parameters generate higher peaks and longer tails in the traffic popularity distribution.

\section{Numerical Results}\label{sec: numerical results}
\begin{figure}
\centering
\includegraphics[width=0.9\columnwidth]{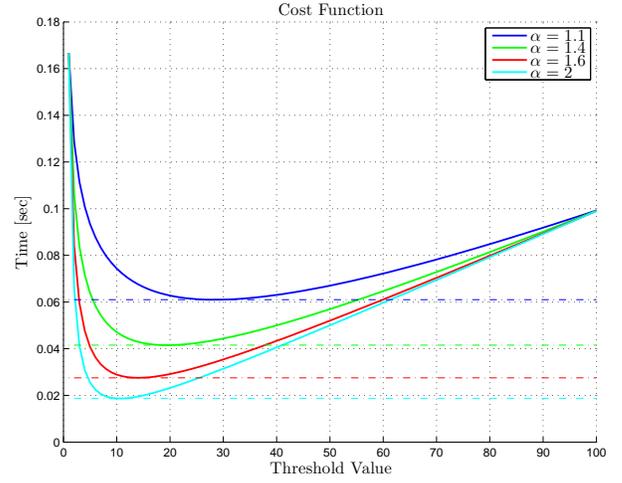}\\
\caption{Cost function versus a varying threshold value \(i_{th}\), for various traffic parameters \(\alpha\). The minimum value of each function is also pointed out. }\label{fig:  cost}
\end{figure}

\begin{figure}
\centering
\includegraphics[width=0.9\columnwidth]{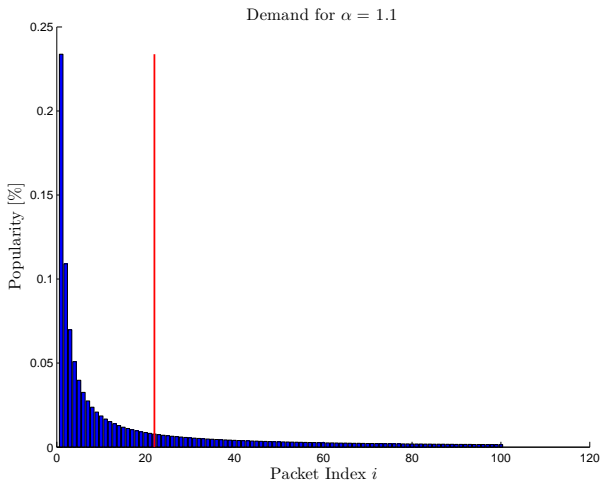}\\
\caption{Traffic demand versus packet/file index, for a traffic parameter $\alpha = 1.1   $ and optimal threshold value between broadcasting and unicasting data. }\label{fig:  demand}
\end{figure}

To the end of establishing two representative values for $spf_\mathrm{bc}$ and $spf_\mathrm{uni}$ let us consider a linear arrangement of BSs, according to the well-known Wyner model \cite{Wyner1994}. Over these cells, random uniformly distributed users are generated. Signals propagate following an exponential path loss model  with exponent $ \eta= 3$.  Rayleigh fading is assumed on top of the path loss model. In more detail a channel coefficient between the $i$-th user and the $j$-th BS is generated as
\begin{align}\label{eq: channel}
h_{ij} = \frac{g_{ij}}{1+\left(d_{ij}\right)^{\eta/2}}
\end{align}
where $g_{ij}\backsim\mathcal{C}\mathcal{N}(0,1)$ and $d_{ij}$ is the distance between the $i$-th user and the $j$-th BS. By normalizing the noise power to one, the power of \eqref{eq: max-min fair power const1} and \eqref{eq: max-min fair power const2} represents the transmit signal to noise ratio (SNR).
In this setup, MC simulations are performed to deduce an average spectral efficiency of the two operations, for $N = 6$  BSs, \(K=500\) system users, \(i_{max}=100\) files and \(P=1\) dBW per-BS power. In broadcast mode, the precoders are calculated by the optimization problem $\mathcal{BC}$ . An average spectral efficiency for this system is calculated as  $spf_\mathrm{bc} = 1 $ b/s/Hz. In the unicast mode of operation, the $\mathcal{UC}$ optimization gives an average spectral efficiency of operation of $spf_\mathrm{uc} = 3$ b/s/Hz. Consequently, the ratio of spectral efficiencies is calculated as: $spf_\mathrm{uni}/ spf_\mathrm{bc}=3$.
Based on this setting, the cost function of problem $\mathcal T$ is plotted in Fig. \ref{fig:  cost} for a range of traffic shaping parameters. It should be noted that the two extremes of this figure represent the time needed by only cellular (far left) and only broadcast (far right delivery). For $a=1.1$, which is considered a typical value \cite{erman2011cache}, the data traffic along with the derived optimal point (red line) are plotted in Fig. \ref{fig:  demand}.

\section{Conclusion}
\label{sec: conclusion}
In this paper, a new approach has been proposed for improving the throughput of multimedia delivery in 5G wireless networks by combining the advantages of unicast and broadcast transmissions. Taking into account multimedia content popularity statistics and proactive caching, highly popular content can be efficiently broadcasted with high spectral efficiency and cached by the majority of users for ``virtually" on-demand consumption. On the other hand, rarely requested content is unicasted through the cellular network to the individual users for real-time consumption. In this context, an analytic framework was elaborated for determining the popularity threshold which delimits the files to be broadcasted from the files to be unicasted.
In this direction, numerical results have demonstrated throughput improvements of up to 80\% in comparison to contentional single-mode systems.

\textit{Acknowledgments}:
This work was partially supported by FNR under the project ``SeMIGod".
\bibliographystyle{IEEEtran}
\bibliography{refs/IEEEabrv,refs/referencesICC2015}
\balance
%
\end{document}